# Densification of Single-Walled Carbon Nanotube Films: Mesoscopic Distinct Element Method Simulations and Experimental Validation

Grigorii Drozdov[1], Igor Ostanin[2,3], Hao Xu[4,*], Yuezhou Wang[5], Traian Dumitrică[1,4,6,**], Artem Grebenko[3,7], Alexey P. Tsapenko[8], Yuriy Gladush[3], Georgy Ermolaev[9], Valentyn S. Volkov[9], Sebastian Eibl[10], Ulrich Rüde[10], Albert G. Nasibulin[3,11]

[1]*Scientific Computation Program, University of Minnesota, Twin Cities, MN 55455, USA*

[2]*Multi-Scale Mechanics, Faculty of Engineering Technology, MESA+, University of Twente, 7500 AE Enschede, Netherlands*

[3]*Skolkovo Institute of Science and Technology, Nobel St. 3, Moscow, Russia*

[4]*Department of Aerospace Engineering and Mechanics, University of Minnesota, Twin Cities, MN 55455, USA*

[5]*Integrated Engineering, Minnesota State University, Mankato, MN 56001, USA*

[6]*Department of Mechanical Engineering, University of Minnesota, Twin Cities, MN 55455, USA*

[7]*Moscow Institute of Physics and Technology, Dolgoprudniy, Institute Lane 9, Russia*

[8]*Department of Applied Physics, Aalto University School of Science, FI-00076 Aalto, Finland*

[9]*Center for Photonics and 2D Materials, Moscow Institute of Physics and Technology, Dolgoprudny 141700, Russia*

[9]*Friedrich-Alexander University Erlangen-Nuremberg, Cauerstr.11, Erlangen, 91052, Germany*

[10]*Aalto University, Department of Materials Science and Engineering, 00076, Aalto, Finland*

*Current Address: *Key Laboratory of Mechanical Reliability for Heavy Equipment and Large Structures of Hebei Province, Yanshan University, Qinhuangdao 066004, China*.

**Corresponding Author. Tel. 612-625-3797. E-mail: dtraian@umn.edu

**Abstract**

Nanometer thin single-walled carbon nanotube (CNT) films collected from the aerosol chemical deposition reactors have gathered attention for their promising applications. Densification of these pristine films provides an important way to manipulate the mechanical, electronic, and optical properties. To elucidate the underlying microstructural level restructuring, which is ultimately responsible for the change in properties, we perform large scale vector-based mesoscopic distinct element method simulations in conjunction with electron microscopy and spectroscopic ellipsometry characterization of pristine and densified films by drop-cast volatile liquid processing. Matching the microscopy observations, pristine CNT films with finite thickness are modeled as self-assembled CNT networks comprising entangled dendritic bundles with branches extending down to individual CNTs. Simulations of the film under uniaxial compression uncover an ultra-soft densification regime extending to a ~75% strain, which is likely accessible with the surface tensional forces arising from liquid surface tension during the evaporation. When removing the loads, the pre-compressed samples evolve into homogeneously densified films with thickness values depending on both the pre-compression level and the sample microstructure. The significant reduction in thickness, confirmed by our spectroscopic ellipsometry, is attributed to the underlying structural changes occurring at the 100 nm scale, including the zipping of the thinnest dendritic branches.

**Keywords:** carbon nanotubes, densification, mechanical properties, mesoscale simulations

## 1. Introduction

Carbon nanotubes films exhibit a unique high porosity/low-density network structure with thickness values down to a few nanometers.[1-4] Mechanical, electrical and optical properties of the CNT films depend greatly on its microstructure, with the CNT orientation (aligned or random network) and density emerging as important parameters in addition to CNT length and diameter. While manufacturing well-aligned films requires sophisticated synthesis techniques[5] and post-processing,[6] the manipulation of the film density is a less technologically challenging process. After synthesis, one usually obtains sparse CNT "pristine" networks that can be densified even unintentionally, for example during the sample transferring and doping procedures. In more controlled approaches, films and yarns are subjected to mechanical[7,8] and liquid phase processing,[9,10] where densification is viewed as an enabling step for developing CNT applications. For example, it was shown that the densification of the vertically aligned CNT forests can lead to significant improvements in mechanical properties.[11] It is also known that the CNT densification enables the CNT-CNT connectivity, which is very important for electrical properties. Thus, the manipulation of the intertube crossing points can be used for developing sensors or flexible electronics devices.[12,13] In the optical domain, pristine films demonstrate antireflection properties, which are destroyed by densification.[14,15]

Understanding the densification process at the microstructural level is a necessary step for establishing a structure-property relationship, which opens up the possibility of making more informed structural modifications toward a specific application. One way to address the complexity presented by the microscale is through simulations.[16] Unfortunately, the wide range of time and length scales encountered in CNT films make the established all-atom molecular dynamics methods[17] computationally prohibitive. Currently, the dynamical simulations of CNT films rely on numerical approaches which are still under development and validation. In this respect, a recent theme in the development of numerical model is based on the idea of coarse-graining,[18-21] where the goal is to develop computationally tractable models of reduced complexity, while preserving to a sufficient accuracy a set of target properties. One of the first developed models was based on the simple bead-and-spring (B&S) polymeric chain model,[18] which is usually trained to reproduce elastic and adhesion properties of CNTs. Unfortunately, the impact of the very limited number of training parameters and (hence limited essential target properties passed to the mesoscale) afforded by the B&S model onto the simulated self-assembled CNT networks is not fully understood. This is important because the simplicity and ease of implementation into existing molecular dynamics codes[17] make the B&S model a popular choice for studying CNT systems as well as other semiflexible polymer networks.[22] So far, one significant deficiency of the B&S model associated with the large artificial barriers for relative displacements of interacting CNT,[19,20] which limit the CNT-CNT sliding during the system evolution, was shown to strongly affect bundling of the self-assembled CNT networks. More complex mesoscopic CNT models,[23] including the mesoscopic distinct element method (mDEM)[20,24] employed here, are placing a premium on the accuracy of the coarse-grained non-bonded[25] and bonded interactions, as well as on accounting for the energy dissipation associated to the CNT tribology.[26]

For validation purposes, we have first examined with electron microscopy real pristine and densified films, which reveal multiscale structural features comprising CNT bundles with thickness values down to individual CNTs. In film benchmark simulation, a stable ultra-thin coupon with few-CNT branches and individual CNTs is regained with mDEM, but not with B&S, which gives instead an adhesion-dominated densified cellular structure formed by thick bundles.

Therefore, the availability of the experimental microstructure for the validation step provides the opportunity to uncover another important deficiency of the B&S model, namely its failure to properly capture the entanglement present in films formed by CNTs with low inclination angles and thus to more strongly usher in accurate mesoscopic models.

Once validated, mDEM is used to successfully generate low-density CNT samples, which comprise interacting CNT bundles with a dendritic structure, and to further simulate surface compression-driven densification. We note that while the mechanical response to uniaxial compression of vertically aligned CNT forests has been recently simulated and interpreted using the concept of a phase transformation,[27,28] isotropic CNT thin films were previously considered only under compressive forces directed along the film direction.[29] The goal of this work is to simulate with a mesoscopic modeling method from the accurate class the compression response of the CNT microstructure over a large deformation range in order to shed light into the film densification mechanism by surface tension. Our simulations reveal the important role of the restructuring processes at the 100 nm scale, delineate the film densification limits of the drop-cast volatile liquid technique, and uncover the structural characteristics of the free-standing densified films. This is foundational knowledge for the establishing of the structure-property relationship, which is needed for the development of future technologies based on CNT films.

Our paper is organized as follows. Section 2 gives an overview of the two coarse-grained simulation methods used here, MDEM and B&S. Section 3 presents our electron microscopy results revealing the complexity of the structural organization of the pristine and densified CNT films, as well as results of express film thickness measurements by spectroscopic ellipsometry. The structural characteristics of the simulated samples, dynamical compression, and recovery simulations are presented in Section 4. Finally, the summary of this computational and experimental study is given in Section 5.

## 2. Computational methods for simulating the mesoscale dynamics

### 2.1. B&S for CNTs.

As already mentioned, the simple B&S polymeric chain multiscale model[18] inherits a very limited number of parameters from the atomistic scale. We briefly review this model to help relate the microstructure features observed in the simulation to the mesoscopic model. In B&S, a CNT of length $L_{CNT}$ is represented as a linear chain of point masses. Each mass $m$ is spaced apart by distance $t$. Massless linear springs between bonded pairs ($k_s$) and triples ($k_b$) are trained to capture the linear extension and bending behavior of individual CNTs, regardless of the size of the CNT segment represented by each spring. More specifically, $k_s = YA/t$ and $k_b = YI/t$, where $Y$ is the Youngs modulus, $A$ the cross-sectional area, and $I$ the area moment of inertia of the CNT. Considering successive $i$, $j$, and $k$ beads, the total potential energy is expressed in terms of two $u_{ij}^t$ and three $u_{ijk}^b$ body potential energy terms

$$u_{ij}^t = \frac{1}{2} k_s (r_{ij} - t)^2 \; ; \; u_{ijk}^b = \frac{1}{2} k_b (\theta_{ijk} - \pi)^2 \; , \tag{1}$$

where $r_{ij}$ stands for the distance between beads $i$ and $j$, while $\theta_{ijk}$ is the angle between consecutive beads $i$, $j$, and $k$. Table 1 lists an example parametrization of a (10,10) CNTs, based on $Y$=1,029 GPa value. The dynamics of the CNT represented this way is dictated by the stretching and bending forces derived from these terms, along with the van der Waals forces described next.

**Table 1**. Parametrization of a B&S model with adhesion for a (10,10) CNT.

| $m$ (amu) | $t$ (Å) | $k_s$ (eV/Å) | $k_b$ (eV) | $\epsilon$ (eV) | $\sigma$ (Å) |
|---|---|---|---|---|---|
| 662 | 3.4 | 270.0 | 6,599 | 0.10 | 12.0 |

The van der Waals inter-tube interactions are described by a pair-wise Lennard-Jones (L-J) potential acting between the beads located on different CNTs. The two available parameters energy $\epsilon$ and distance $\sigma$ are usually fitted to describe the spacing and adhesion between two parallel CNTs. For (10,10) CNTs, the adhesion is 0.23 eV/Å. Considering the tubes in contact, the inter-axis distance becomes 13.5 Å, one obtains the parameters listed in Table 1.

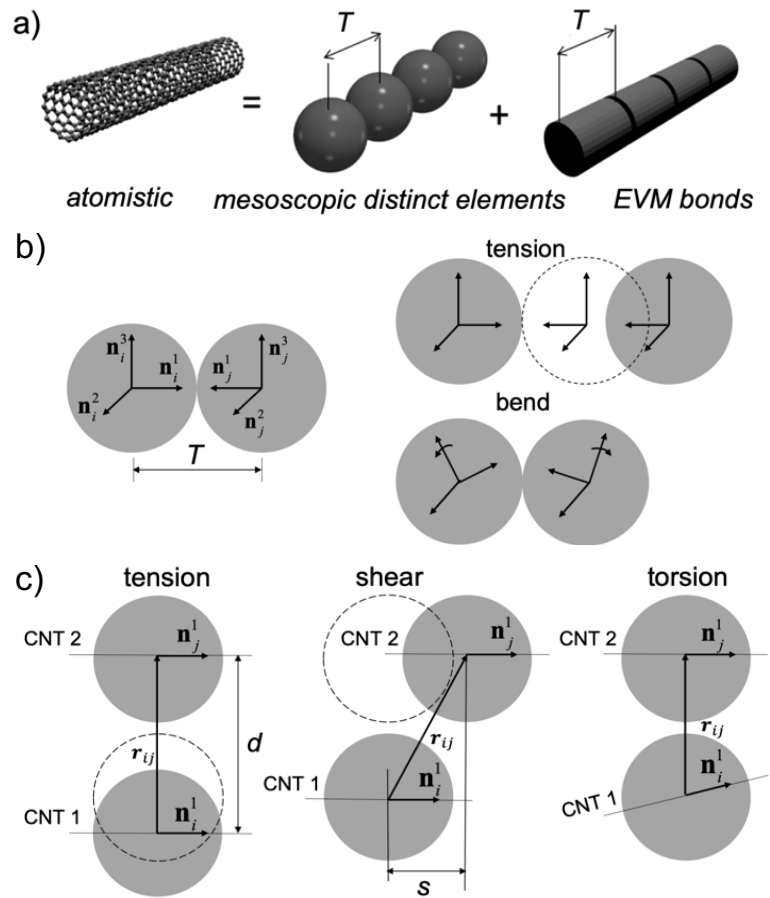

**Figure 1.** a) Coarse graining of the atomistic representation of a CNT into mesoscopic distinct elements connected by vector-based bond models. b) Schematics of two distinct elements linked by an enhanced vector model (EVM) bond (left) and two modes of the bonded deformation (right). c) Three deformation modes of the van der Waals interactions of elements located on different CNTs, for which only the axes are shown. The empty circles illustrate the position of the particles before deformation.

Previous studies[19,20] have shown that B&S chains can represent the shear strength of interacting CNTs when each bead represents only a small segment of a CNT. This is because the isotropic L-J potential between two infinite parallel chains favors a staggered over an in-line arrangement, a behavior that is not seen at the molecular level. Therefore, simulating the CNT dynamics requires the use of "small" beads, like the ones shown in Table 1 derived for a (10,10) CNT, although coarser grains are required to achieve simulations of large enough CNT network samples that exhibit representative behavior.[30]

Finally, it is important to note that the B&S model is a thermal model. After assigning velocities corresponding to the desired temperature, the beads can be evolved in time in the NVT ensemble with molecular dynamics codes, like LAMMPS.[17]

*2.2. mDEM for CNTs*

The mDEM model for CNTs[20,24] was introduced in our previous works in conjunction with its implementation in the popular distinct element code PFC3D.[31] More recently, the model was adapted to a vector-based format[32] and implemented[30,33] in the code WaLBerla.[34] Relying on a scalable message passing interface framework, WaLBerla enables deterministic evolution of millions of distinct elements of given mass and moment of inertia.

Figure 1a illustrates the mDEM representation of a CNT (left), where each mesoscopic distinct element (center) lumps a finite number of atoms of the atomistic model. The distinct elements in direct contact interact via the enhanced vector model (EVM) potential,[35] which describes the resistance to stretching, shear, bending, and torsional deformation displacements of the CNT portion of length $T$ and radius $r_{CNT}$ as a Euler-Bernoulli beam. Consider two equal-sized particles $i$ and $j$ with equilibrium separation $T$ and orientation described with orthogonal vectors $\mathbf{n}_i^k$ and $\mathbf{n}_j^k, k = 1,2,3.$ attached to each particle. The interaction potential is given as the function of invariant quantities associated with the inter-center normal $\mathbf{e}_{ij}$ and two set of vectors $\mathbf{n}_i^k, \mathbf{n}_j^k$. Let $\mathbf{n}_i^1$ be along the tube long axis. As depicted in Fig. 1b (left), in the undeformed state $\mathbf{n}_i^1 = -\mathbf{n}_j^1$, $\mathbf{n}_i^2 = \mathbf{n}_j^2$, and $\mathbf{n}_i^3 = \mathbf{n}_j^3$. The key components (for the present investigation) of the two-body EVM potential $U_{ij}$ describe the resistance to stretching $U_{ij}^t$ and bending $U_{ij}^b$ deformation of the two elements, Fig. 1b right, as [33,35]

$$U_{ij}^t = \frac{YA}{2T}(r_{ij} - T)^2 \; ; \; U_{ij}^b = \frac{YI}{T}(\mathbf{n}_i^1 \cdot \mathbf{n}_j^1 + 1) \qquad (2)$$

where $Y$ is the CNT Young modulus, $A$ is the CNT cross-section area, and $I$ the area moment of inertia. EVM replaces the parallel-bond incremental model, employed in the earlier model.[20]

Unlike in B&S, the mDEM dynamics is dictated not only by forces but also from moments arising from the employed vector-based potentials. For example, the components of force $\mathbf{F}_{ij}$ and moment $\mathbf{M}_{ij}$ on particle $i$ due to particle $j$ corresponding to terms (2) write down as

$$\mathbf{F}_{ij} = \frac{YA}{2T}(r_{ij} - T)\mathbf{n}_r \; , \; \mathbf{M}_{ij} = \frac{YI}{T}\mathbf{n}_i^1 \cdot \mathbf{n}_j^1 \qquad (3)$$

Here $\mathbf{n}_r = \mathbf{r}_{ij}/r_{ij}$. In this work, we coarse-grained a (10,10) CNT into a chain of spherical distinct elements with diameters $2r_{CNT}$ corresponding to CNT portions with $T = 2r_{CNT} = 13.56$ Å, i.e., four times larger than the in the B&S model shown in Table 1. The values of the parameters of equation (2) are $Y = 1,029$ GPa, $A = 142.7$ Å$^2$, and $I = 3,480$ Å$^4$.

The coarser mDEM modeling of CNTs is also enabled by a two-body vdW potential, which nevertheless is of a more sophisticated form than the simple parameterized L-J potential. The non-bonded interaction potential between elements takes a symmetrized form:

$$U_{ij} = U(r_{ij}, \mathbf{n}_r \cdot (\mathbf{n}_i^1 + \mathbf{n}_j^1), \mathbf{n}_i^1 \cdot \mathbf{n}_j^1) = U(r_{ij}, \cos\theta_{ij}, \cos\gamma_{ij}) \quad (4)$$

where the angles $\theta_{ij}$ and $\gamma_{ij}$ describe mutual orientation of two elements. The precise expression for the potential is presented in ref. [8]. It should be noted that this form captures not only normal forces, but also shear forces and aligning moments between distinct elements, Fig. 1c.

$$\mathbf{F}_{ij}^r = -\frac{\partial U}{\partial r_{ij}} \mathbf{n}_r \quad (5a)$$

$$\mathbf{F}_{ij}^\theta = -\frac{1}{r_{ij}} \frac{\partial U}{\partial \theta_{ij}} \mathbf{n}_\theta, \quad \mathbf{n}_\theta = \frac{\mathbf{r}_{ij} \times (\mathbf{r}_{ij} \times (\mathbf{n}_i^1 + \mathbf{n}_j^1))}{\left| \mathbf{r}_{ij} \times (\mathbf{r}_{ij} \times (\mathbf{n}_i^1 + \mathbf{n}_j^1)) \right|} \quad (5b)$$

$$\mathbf{M}_{ij}^\gamma = -\frac{\partial U}{\partial \gamma_{ij}} \mathbf{n}_\gamma, \quad \mathbf{n}_\gamma = \frac{\mathbf{n}_i^1 \times \mathbf{n}_j^1}{\left| \mathbf{n}_i^1 \times \mathbf{n}_j^1 \right|} \quad (5c)$$

The partial derivatives $\frac{\partial U}{\partial r_{ij}}, \frac{\partial U}{\partial \theta_{ij}}, \frac{\partial U}{\partial \gamma_{ij}}$ are given in Ref. [8] and are therefore not repeated here.

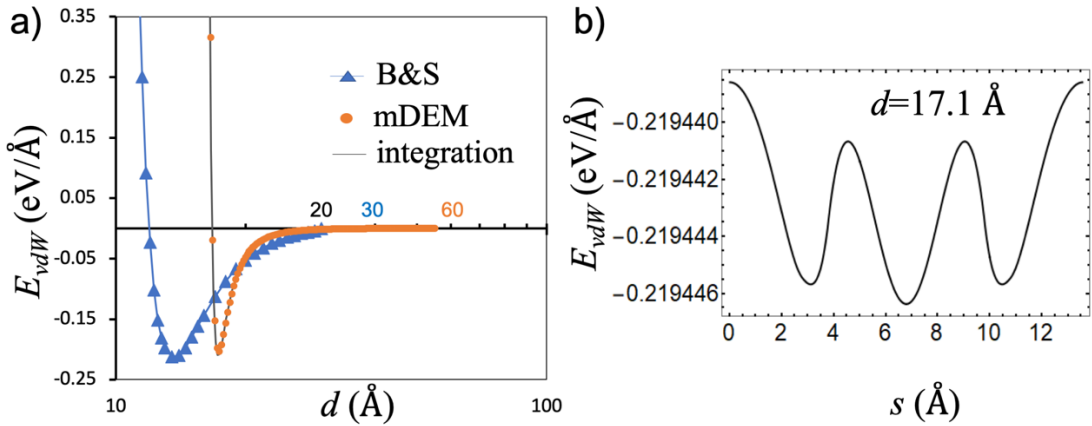

**Figure 2.** a) The vdW energy $E_{vdW}$ (per unit length) versus $s$ the tube-tube center-to-center distance $d$, as described with mDEM (circles), B&S (triangles connected by lines), and direct integration (continuous line). b) The mDEM model for non-bonded shear deformation of two distinct elements (left) leads to a constant $E_{vdW}$ vs $s$, the relative shift between the two parallel infinitely long CNTs at $d$ =17.1 Å, where the potential depth has a minimum.

To demonstrate this potential, Figure 2a plots the vdW potential energy between two parallel (10,10) CNTs, each represented by 10 distinct elements placed under periodic boundary conditions (PBC). It can be seen that the mDEM model describes precisely the narrow well obtained by direct integration of the atomistic L-J interactions.[20,36] At $d =17.1$ Å, $E_{vdW}$ has a minimum, where the force between CNTs is zero. At smaller $d$, the potential increases very rapidly and becomes positive for $d <16.8$ Å. This behavior corresponds to significant repulsive forces between the elements located on the two CNTs, which prevent the interacting elements to interpenetrate. At larger $d$, the attractive dispersive interactions are dominating, and this will result in clumping of the neighboring CNT's.

While the potential well description based on the simpler L-J potential of the B&S model captures similar adhesion energy, there are also some pronounced differences with the integrated well. By having the minima at $d =13.5$ Å, the B&S allows for the CNT surfaces to physically overlap and even interpenetrate. On the attractive side, the potential approaches the horizontal axis with a lower curvature, and even intersects the attractive part of the true integrated potential at $d$ ~20 Å. Thus, we expect that in the B&S model, the corresponding vdW attractive force between CNTs will have a much larger effective range than the underlying atomistic model.

Figure 2b illustrates another important quality of the mDEM model. Through the dependence of the non-bonding potential on the angle made by the $\mathbf{n}^1_i + \mathbf{n}^1_j$ and $\mathbf{r}_{ij}$ vectors, i.e., the shear deformation in Figure 1d, $E_{vdW}$ is made invariant to relative shifts $s$ of parallel CNTs separated by the maximal adhesion distance. Thus, despite the large $T$ of the distinct elements, the mDEM inter-tube interaction is free of the artificial corrugation introduced by the B&S type of mesoscopic CNT models, and therefore provides realistic smooth sliding of aligned CNTs.

The *athermal* model overviewed above is augmented to capture the dissipative atomistic scale processes associated with CNT sliding against one another.[24,26,37] Specifically, mDEM presents two channels for energy dissipation: a viscous damping and a local damping. The effective dissipative force acting between two aligned CNTs is captured in mDEM by dashpot viscous contacts with constitutive equation $\mathbf{F}_d = \eta \mathbf{v}_r$ acting in parallel with the vdW contacts. Here $\mathbf{F}_d$ is the dynamic friction force, $\mathbf{v}_r$ the relative velocity of the distinct elements in vdW contact, and $\eta$ is the linear viscous coefficient. Local damping adds forces ($\mathbf{F}_\alpha$) and moments ($\mathbf{M}_\alpha$), $\mathbf{F}_\alpha = -\alpha \mathbf{F} \, sign(\mathbf{v})$ and $\mathbf{M}_\alpha = -\alpha \mathbf{M} \, sign(\mathbf{\Omega})$ where $\mathbf{v}$ and $\mathbf{\Omega}$ are the translational and rotational velocities of an element, respectively, and $\alpha$ is the local damping parameter. Local damping is introduced in the classical DEM[31] with the sole goal of stabilizing the numerical time integration. The values $\alpha = 0.4$ and $\eta = 0.12$ pN s/m used here were selected to match the relaxation of two crossed (10,10) CNTs and the dynamical friction between two aligned (10,10) CNTs.[26]

The representation of the bonded and non-bonded interactions with potentials allows for symplectic integration of the equation of motion instead of the incremental algorithm used for parallel bonds in PFC3D.[31] More specifically, the dynamics of distinct elements is described in terms of state variables – translational and rotational positions and velocities of distinct elements – that are evolved in time with an explicit velocity Verlet time integration scheme. The evolution of rotational degrees of freedom is analogous with the rotations stored as quaternions.[38]

### 3. Experimental Methods and Characterization of Pristine and Densified Films

To infer the structure of CNT films, we have first investigated with both SEM and TEM high-quality films produced by a floating catalyst (aerosol) chemical deposition.[39,40] These "pristine" films comprise *μ*m-long 1.4−2.3 nm in diameter single-walled CNTs, which are randomly oriented in the film's plane as they are collected onto cellulose membrane filters directly from the reactor. They have 10−200 nm in thickness and densities of ~0.12 g/cm³. In addition, the precise film thickness before and after densification was measured with spectroscopic ellipsometry.

*3.1. Electron Microscopy Results*

The combined SEM and TEM investigations evidence a hierarchical bundling pattern in the pristine CNT films. A typical few μm portion of the film surface can be observed in the SEM image presented in Figure 3a. The structure consists of wavy CNT bundles, which are branching out into thinner and bent bundles or merging into larger ones. On the 100 nm scale, Figure 3b, thinner bundles comprising few-CNTs and even individual CNTs can be observed with TEM. Figure 3c, details few-CNT bundles, some branching out in fewer-CNT bundles and individual CNTs, coexisting with thicker ~10 nm diameter "medium" bundles. The TEM images show only the projection of the bundle on the 2-dimensional plane and do not have information about the 3-dimensionallity of the structure. As such, the number of CNTs in these "medium" bundles can only be estimated based on the measured bundle thickness and the mean diameter $2r_{CNT}$ of individual CNTs. For example, a ~10 nm bundle with a hexagonal closed-packed $2r_{CNT}=2.2$ nm single-walled CNTs [41] would contain about 14 CNTs. Figure 3d, details the zipping of three "medium" bundles into a larger bundle. As in the previous images, individual CNTs and thin 2-CNT bundles are also present. Interestingly, these bundles and others are oriented perpendicularly onto the 3 "medium" horizontal bundles indicated by arrows. This orientation suggests that these bundles could play a role in arresting the zipping progression of the 3 "medium" bundles.

Significant densification can be achieved by drop-cast volatile liquid processing.[41] As it can be seen in the SEM image shown in Figure 3e, on the few μm scale such densified films preserve the branching characteristic observed earlier in pristine films. However, zooming in to the 100 nm scale, Figure 3f, the TEM images no longer show fewer-CNT bundles and individual CNT features. Additionally, this image shows the zipping (but not the full coalescence) the "medium" bundles located at different heights.

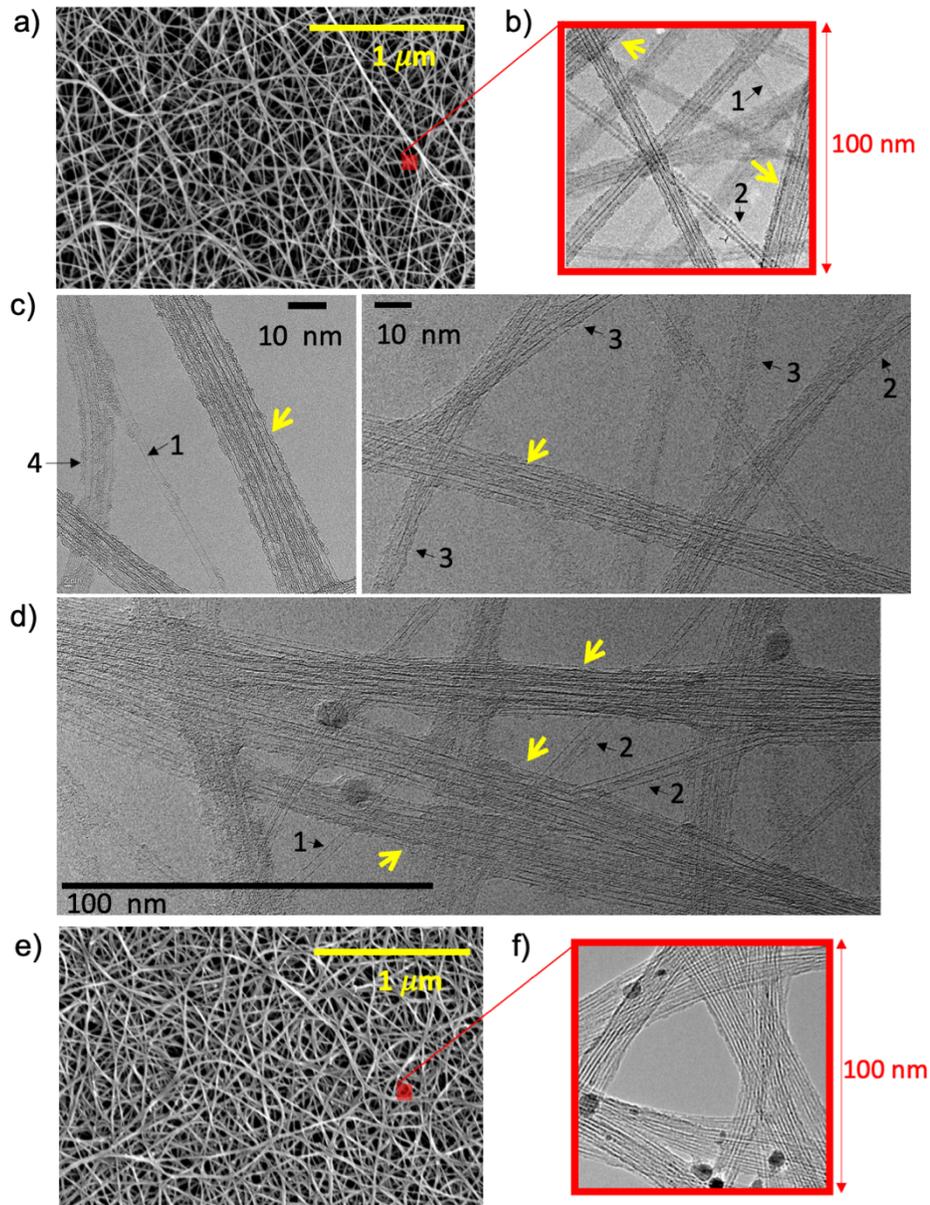

**Figure 3.** Multiscale structure of the pristine a)-d) and densified e)-f) CNT film: a) SEM showing an entangled structure of µm-long wavy bundles. b) TEM of individual and 2-CNT bundles coexisting with "medium" ~10 nm diameter bundles (indicated with yellow arrow heads) on a 100 nm scale. TEM detailing c) ~10 nm diameter bundles and d) their merging into a thicker "large" bundle. Individual and 2-3 CNT bundles (indicated by black arrows) can be still observed. e) SEM showing µm-long wavy bundles. f) TEM of few-CNT bundles zipped on a 100 nm scale.

*3.2. Spectroscopic Ellipsometry Results*

To characterize quantitatively the change in thickness under densification, we produced a free-standing CNT film by dry-transfer method and measured its thickness before and after its densification by drop-casting of 200 μl clean ethanol. We use spectroscopic ellipsometry, a method that ensures that the film morphology is not affected by the measurement. For the measurement on the densified film we have followed the methodology described in our recent work[3] where we successfully demonstrated that reflectance spectroscopic ellipsometry could alone determine the thickness and optical constants of the freestanding densified CNT films. For pristine film we had to modify the method to measure in transmission (Figure 4a) because the film reflection is suppressed to less than 1%. For both types of measurement, we use a variable-angle spectroscopic ellipsometer (VASE, J.A. Woollam Co.) within the wavelength range 300 – 600 nm.

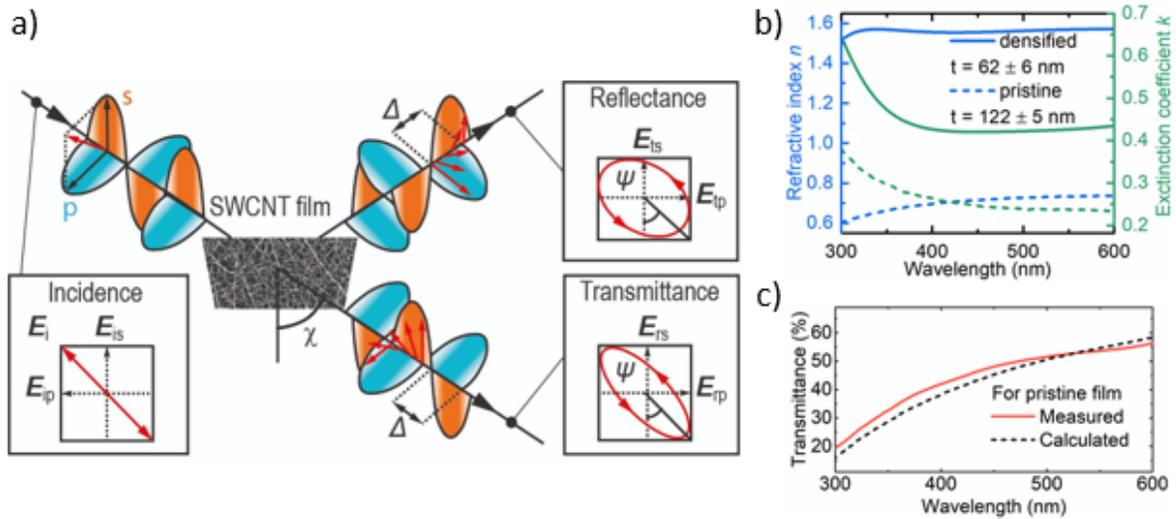

**Figure 4.** a) Scheme of ellipsometry measurement. b) Obtained optical constants – refractive index (blue) and extinction coefficient (green), for pristine (dashed line) and densified (continuous line) CNT films. c) Comparison of the transmittance spectra as obtained from the measurement (continuous line) and calculated (dashed line) from the measured optical constants.

We have strong indications that pristine films exhibit optical anisotropy, which makes the simultaneous determination of optical constants and thickness a complicated task. However, as was shown by Hilfiker and co-workers,[42] the thickness could be determined from standard ellipsometry fitting of the change in polarization (i.e., Δ and Ψ in Figure 4a) with effective optical constants. To verify the approach we have retrieved the optical constants, i.e., the refractive index $n$ and the extinction coefficient $k$ as plotted in Figure 4b, and then calculated the transmission predicted by the transfer matrix method.[43] We obtained an excellent agreement with experimental values, Figure 4c. Employing these two approaches for a given film, we determined a thickness of $h_o$=122±5 nm for the pristine case and a $h_d$=62±6 nm thickness after densification. This means the densification reduces thickness by about 50%.

### 4. Simulation Results

*4.1. Spontaneous densification of an ultrathin "pristine" sample: mDEM vs. B&S*

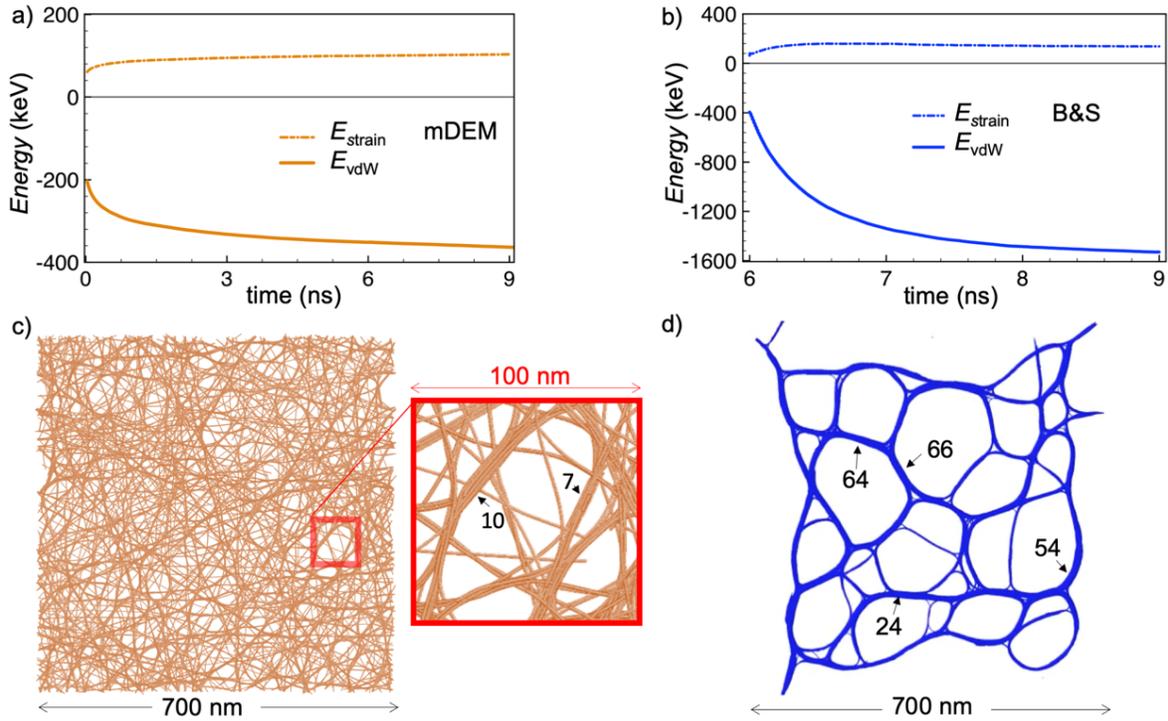

**Figure 5.** Comparison of the vdW and bending energy evolution in a) mDEM and b) B&S representation of a film network 700 nm x 700 nm x 11 nm in size and density of $\rho = 0.176$ g/cm$^3$. The film morphologies (top view) at the end of relaxation with c) mDEM and d) B&S; The number of CNTs in the selected bundles indicated by arrows is also given.

Focusing now on simulations, we have first tested the ability of the mDEM and B&S models to describe the microstructure observed on the sub-*μ*m scale of the low-density pristine CNT films. In a benchmark example, Figure 5, we find that the mDEM and B&S models evolve into dramatically different structures. We have considered a network containing 450 CNTs, each with $L_{CNT}$=651 nm. The starting structure for the mDEM simulations is a computer-generated network containing straight (10,10) CNTs, "grown" from "seeds" randomly placed in a 700 nm x 700 nm x 11 nm cuboid. The CNT orientations were distributed isotopically in the CNT xy plane, and uniformly distributed out-of-plane within the angle φ made by $\mathbf{n}_i^1$ vectors with the xy plane within the borders of 0 and $11^0$.

The initial van der Waals energy ($E_{vdW}$) of the sample originates in the crossed CNTs that form junctions. During cycling with a 20 fs time step, the evolution is driven by the attractive vdW interactions which act to lower $E_{vdW}$ at the expense of introducing strain energy ($E_{strain}$) in individual CNTs, Figure 4a, which practically coincides with the bending strain. The relaxation process is particularly fast during the first ns of the evolution. At the microstructure level, relaxation occurs through zipping,[13,19] a topological transformation in which crossed CNTs bind locally on a sub-ns time scale.[26] The bundle aggregates generated this way contain only a few CNTs. After ~3 ns, the $E_{vdW}$ lowering slows down significantly as the repulsive components of the vdW interactions and the CNT strain resistance are hindering the coarsening of the network into bundles. This "steady-state" structure contains "medium" size 5-10 CNT bundles, coexisting with thinner bundles and even individual CNTs, Figure 5b. As it can be seen in the inset, the 100 nm scale of the simulated sample exhibit important similarities, in terms of the variety of bundle sizes and branching, with the structure observed in the TEM images of the experimental film.

As ratio between elastic and adhesion energy is well captured by mDEM and B&S, one may expect that both models will give qualitatively similar network structures. In order to compare the outcomes of the mDEM and B&S treatments without introducing any sample bias, we have alternatively evolved the network structure obtained with mDEM after 6 ns of simulation time with the B&S treatment coupled with a Nose-Hoover thermostat at 300K. When switched to the B&S model, the network structure becomes highly unstable. As it can be seen from Figure 5b, $E_{vdW}$ instantly departs from the mDEM value and, during only 3 ns of evolution, it reaches a value which is about four times lower. $E_{strain}$ presents a similar departure from its DEM value, reaching after 3 ns an about 30% larger value.

In both mDEM and B&S simulations, the network evolution is driven by the adhesive vdW forces. As CNT bending is well captured by the two models, it follows that the differences revealed in Figure 5 originate into the inherent deficiencies of the L-J description of the non-bonded interactions, Figure 3a. The B&S model fails to sustain the correct balance between the repulsive volume and dispersion interactions and the "input" CNT network disentangles into a structure dominated by adhesion. During the B&S 3 ns of evolution, CNTs clump into bundles that further zip into very thick 50-70 CNT bundles, leaving behind large empty pores. The film acquires a qualitatively different cellular structure, with CNTs characterized by $\varphi \sim 0$. Such a structural organization of the sub-$\mu$m scale disagrees with the experimental observations of Fig. 3b-d, and is closer to the images of a few-$\mu$m scale.

Beyond the 3 ns time, we expect that the structures will further evolve into individual bundles weakly interacting with each other through bent CNT located at the bundle connectivity points.[19]

*4.2. mDEM simulations of spontaneous densification of "pristine" samples*

The above results make mDEM the method of choice for studying film densification. Although mDEM has the necessary accuracy, developing faithful computational models for the CNT films described in Section 3 is still challenging. In addition to the difficulties associated with identifying the three-dimensionality of the film structure from the SEM and TEM images, we recognize that the multiscale structure observed by electron microscopy is the result of the vdW-driven CNT coagulation during synthesis, which involves direct CNT-CNT collisions in the gas phase,[21] and during and after the collection onto the filter substrate. While the former involves zipping of individual CNTs, the later stage is likely dominated by the bundle-bundle interaction processes.[44] Additionally, there are computational challenges related to the system size.

Understanding the film densification requires a straight representation of the film thickness. In the in-plane direction, the sub-$\mu$m scale is reasonably accessible computationally with the current implementation of mDEM, while the few-micrometer scale still poses extreme computational demands.[30] Without sacrificing the realistic representation of film thickness, we have focused our study on the sub-$\mu$m scale, where the accurate description of the mesoscopic interactions is able to describe individual and few-CNT bundles coexisting with "medium" size bundles. As suggested by the TEM images of Figure 3b and f, this scale undergoes significant restructuring during compression and therefore could be important for the densification process.

**Table 2**. Parameters and statistical information of five CNT samples with initial density of 0.12 g/cm$^3$. The magnitude of the Herman orientation factor <HOF>, average tilt <$\varphi$> of elements, film thickness $h_o$, density $\rho_o$, van der Waals $E_{vdW}$, strain energy $E_{strain}$, and total energy $E$ are listed for the sample structure generated at the end of the relaxation.

| Film & Size ($\mu$m) | $N_{CNT}$ | $L_{CNT}$ ($\mu$m) | $\varphi_{max}$ (°) | <HOF> | <$\varphi$> (°) | $h_o$ (nm) | $\rho_o$ (g/cm$^3$) | $E_{vdW}$ (keV/CNT) | $E_{strain}$ (keV/CNT) | $E$ (keV/CNT) |
|---|---|---|---|---|---|---|---|---|---|---|
| FA 1x1x0.14 | 5185 | 1 | 11 | -0.48 | 4.4 | 140 | 0.12 | -0.91 | 0.21 | -0.7 |
| FB 1x1x0.14 | 5185 | 1 | 11 | -0.49 | 3.8 | 108 | 0.15 | -2.08 | 0.18 | -1.9 |
| FC 0.33x0.33x0.14 | 1733 | 0.33 | 11 | -0.48 | 4.9 | 127 | 0.13 | -0.34 | 0.08 | -0.26 |
| FD 0.33x0.33x0.14 | 1733 | 0.33 | 22 | -0.46 | 7.1 | 133 | 0.13 | -0.27 | 0.07 | -0.20 |
| FF 0.33x0.33x0.14 | 1733 | 0.33 | 45 | -0.35 | 15.4 | 209 | 0.08 | -0.19 | 0.04 | -0.15 |

To this end, we have constructed a series of morphologically-diverse models for the "pristine" CNT film, as summarized in Table 2, with the goal of understanding how the different parameters affect the film structure and densification, and to eventually project this understanding to the behavior of the larger scale. The sample films were generated in the same manner as in the example of Figure 5, and all correspond to a low mass density of 0.12 g/cm$^3$. With the exception of "FB", all the sample models were constructed by "growing" individual CNTs at random locations inside the control volume listed in the first column of Table 2. To get insight into the interaction of already-formed bundles collected onto the filter substrate, in "FB", 2-CNT pairs rather than individual CNTs were grown.

All CNTs considered here are of (10,10) type, which is a reasonable choice for our experimental films. The $L_{CNT}$ values, listed in the 3$^{rd}$ column of Table 2, were selected in order to generate network structures with different degrees of cohesive energies.[26,45] In experiment, the collected CNTs are expected to have low inclinations with respect to the filter plane. Therefore, the constructed samples "FA", "FB", and "FC" used $\varphi_{max}=11°$. In "FD" and "FF", larger $\varphi_{max}$ values were considered in order to explore the impact of the inclination angle parameter, which, along $L_{CNT}$, are expected to couple to the network entanglement. All samples were subjected to PBC along the x and y Cartesian coordinates. As such, the simulated system exhibits two free surfaces, which are perpendicularly-oriented onto the z Cartesian coordinate. Although the boundary conditions make it possible to represent films extending much further than the actual lateral dimensions of the computational system, they do not capture the patterns observed at the few *μ*m scale, which consists of the self-similar branching realized with thicker CNT bundles.

To conduct parallelized simulations, the film domain is divided into cuboid subdomains, which are then distributed between computational cores and connected through the message passing interface. For each timestep, every computational core resolves the portion of contacts associated with the elements that are inside the assigned subdomain. After each integration step, all processes synchronize their data, and distinct elements with updated positions can migrate from one subdomain to another. Although the selected domains are balanced throughout the simulations, the number of contacts is increasing with the film densification and the numerical cycling slows down. Nevertheless, this parallelized framework allowed us to reach efficient simulation times for unprecedented CNT film sizes: for the most computationally intensive simulations, such as compression of the "FA" and "FB" CNT films, where 3.8 million elements interact through up to 300 million contacts, it took around 3 days of real-time when the simulation domain was distributed over 2240 processes on average. Simulations were performed on three computational clusters: Pleiades (NASA's Advanced Supercomputing Division), Zhores (Skoltech University),[46] and Mesabi (Minnesota Supercomputer Institute).

Figures 6a and b displays images of the "FA" and "FB" film, as initially generated, and after 60 ns of mDEM cycling with a 20 fs time step, while Figures 6c and d display the evolution during this process of $E_{vdW}$ and $E_{strain}$, respectively. As in the benchmark example, $E_{vdW}$ ($E_{strain}$) decreases (increases) fast during the first few ns, and continues on at a much slower rate. As before, this behavior corresponds to zipping, a process that also bends the initially straight CNTs.

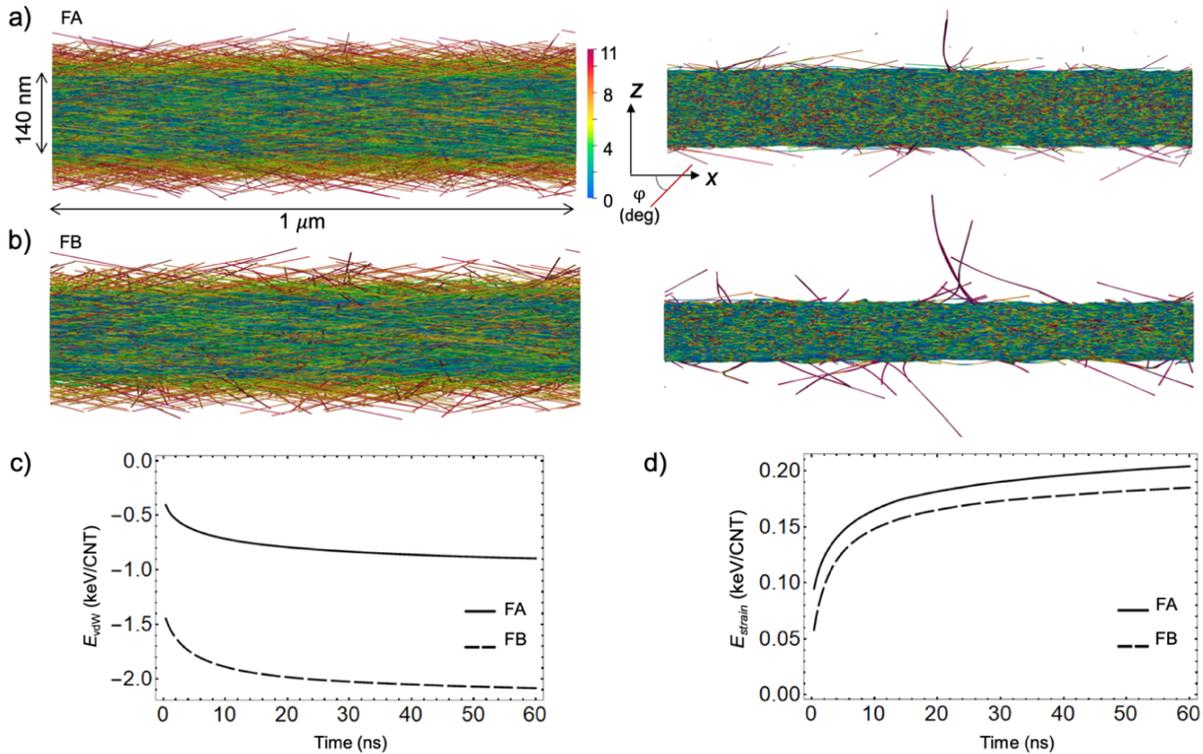

**Figure 6.** mDEM relaxation simulations: The sample films a) "FA" and b) "FB" as computer-generated (left) and after 60 ns of mDEM relaxation (right). The color codes the magnitude of $\varphi$ for each distinct element. The evolution of the c) van der Waals and d) strain energy.

Because zipping occurs in the vicinity of the of junction points, this mechanism is prevalent in the "bulk" part of the film, where the density of distinct elements is largest. Also, because of the small angle of inclination ($\varphi_{max}=11°$) the emerging bundles will be preferentially oriented in the film plane. Regarding the two film surfaces, we note the computer-generated samples have rough surfaces. Many distinct elements on the surface belong to CNTs that are near the maximum inclination angle $\varphi_{max}$. During relaxation, most CNTs become nearly horizontal and zip with "bulk" part of the film, thus creating relatively smooth surfaces. Because of the poor connectivity in the surface region, zipping of these CNTs among themselves is a rare event. When it occurs, surface bundles with a more vertical orientation are formed, as visible in Figures 6a and b.

Focusing now on the bulk part of the sample, Figure 7a shows a typical dynamic leading to "medium" bundles. The two simulation snapshots show three groups of initially crossed individual CNTs, which are located (for better visibility) just under the top film surface. These CNTs are subsequently zipping into bundles containing 6 and 7 CNTs. The top view presented in Figure 7b confirms that during relaxation, these bundles are branching out into thinner bundles and individual CNTs. The branches are sustained through their interaction with the film structure. For example, Figure 7c presents an entanglement mechanism in which the CNTs indicated with black arrows are filling the junction formed by two other CNTs (shown in green and maroon, respectively), thus helping prevent their zipping into bundles. In another mechanism, the purple arrow points to a CNT (shown in maroon) adhered to a side bundle. Such adhesion forces are helping prevent the zipping of this branch to the main maroon bundle. Thus, Figure 7b proposes to conceptualize the film microstructure as a collection of entangled dendrimers, which are interacting through their branches. Without the repulsive volume interaction, van der Waals attractions, and dissipative frictional forces, the dendritic branches would quickly restructure into the main bundle.

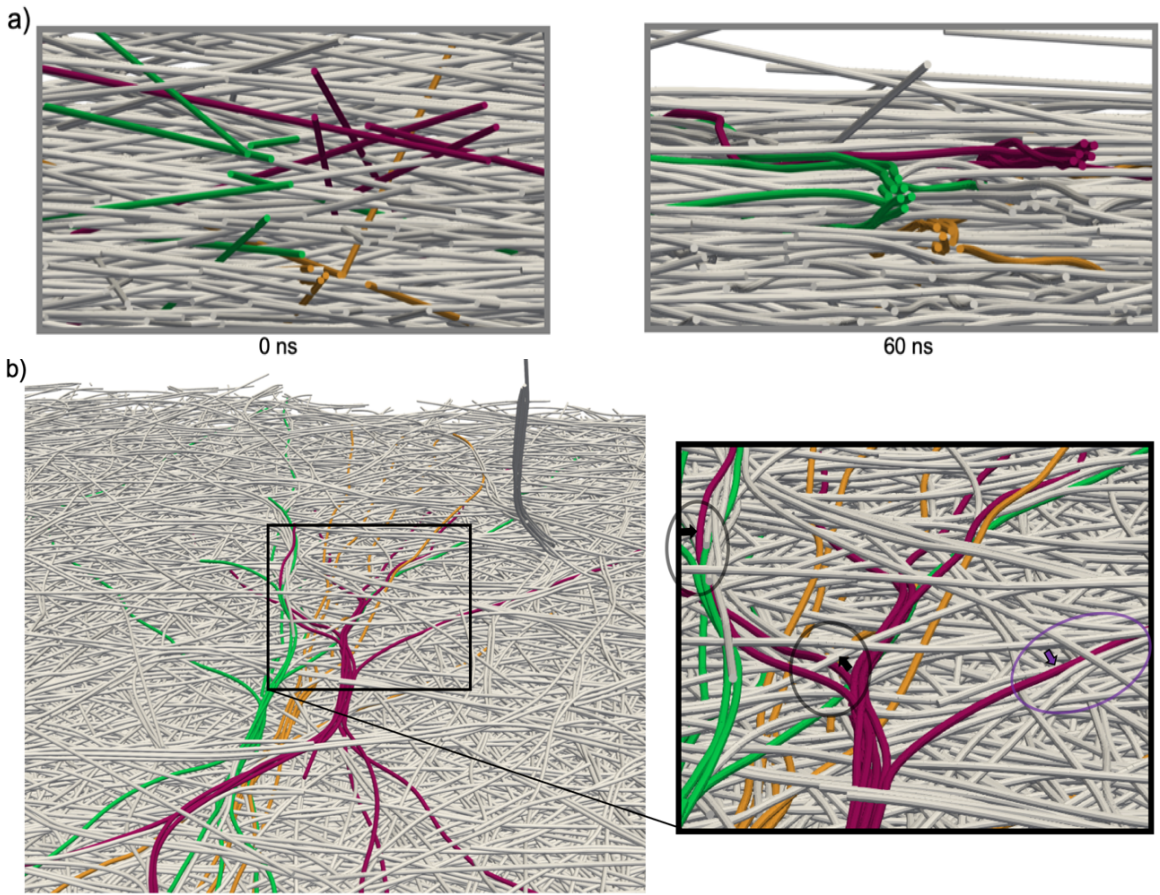

**Figure 7.** mDEM relaxation simulations: a) A region located near the top surface of sample "FA" before (left) and after 60 ns of relaxation (right). The color (green, orange, maroon) is used to distinguish the CNTs that zip into the 3 bundles (green, orange, maroon). b) Birds' eye view of the top surface showing the branching structure of the bundles. The callout shows the arrest of zipping of the branches to the main one caused by excluded volume interactions (black ovals and arrows) and by zipping of these branches to other bundles (purple oval and arrow).

Turning back to Figure 6c, it is important to clarify that while the "FB" sample exhibits a much lower $E_{vdW}$ than "FA", this difference does not imply the onset of an adhesion dominated structure. Inspection of the microstructure showed that as "FA", "FB" exhibits a similar branching pattern, but with a 2-CNT bundle cutoff. The ~1 keV/CNT van der Waals energy advantage of the "FB" sample can be entirely attributed to the initial intra-bundle adhesion introduced by the initial construction and, in fact, "FA" and "FB" exhibit similar inter-bundle cohesive energy values. The $E_{strain}$ of "FB", Figure 6d, is comparable to and slightly lower than the value found in "FA" because the 2-CNT bundles bend independently with inter-tube sliding. Bending with inter-tube sliding is a soft deformation mode captured[47] by mDEM in which the bundle bending rigidity scales with the number of constituent CNTs.

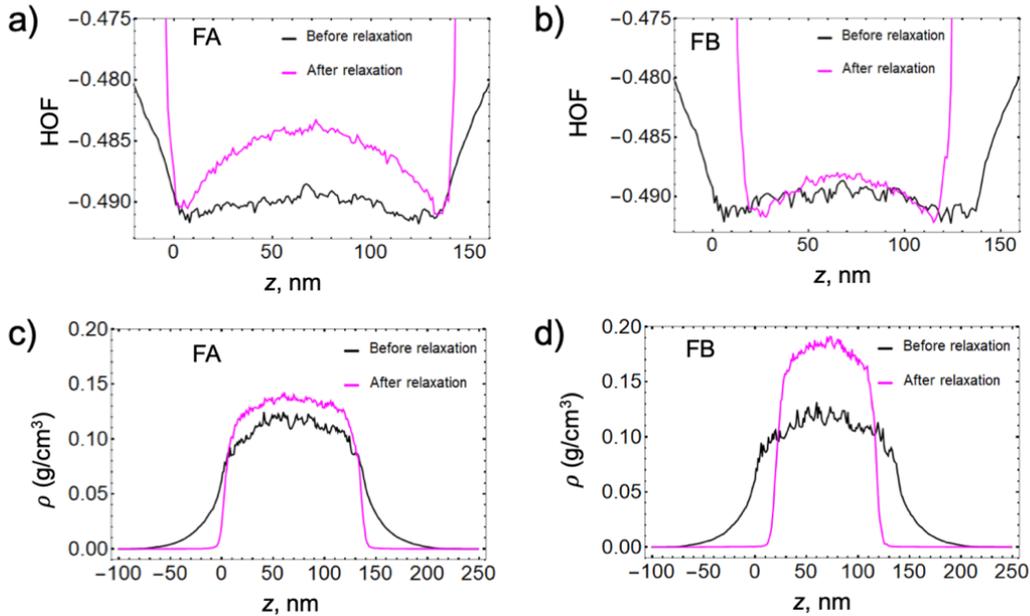

**Figure 8.** mDEM relaxation simulations: Herman orientation factor (HOF) across z in a) "FA" and b) "FB" samples before and after relaxation. Density $\rho$ across z in c) the "FA" and d) "FB" sample before and after relaxation.

As the pristine film thickness is an experimentally accessible parameter, we have monitored the evolution of this parameter in the simulated samples. Starting from the computer-generated networks, we noted that the zipping couples to a fast decrease in CNT film thickness. This is because bundling occurs mainly in the film direction. Nevertheless, as zipping slows down, the film thinning effect becomes negligible and the film acquires a finite thickness. In Figures 6a and b, elements are represented in color according to their orientation Nevertheless, as zipping slows down, the film thinning effect becomes negligible and the film acquires a finite thickness. In Figures 6a and b, elements are represented in color according to their many elements located in the bulk region remain inclined ($\varphi>0$) after relaxation. Based on this observation, we associate the initial arrest of the film thinning to the entanglement developed by inclined CNTs.

For a quantitative measure of the CNT orientation in the film, we have plotted in Figures 8a and b the magnitude of the Herman orientation factor (HOF), defined as $\text{HOF}=\frac{1}{2}[3<\cos^2(90^0-\varphi)>-1]$. Here, the angle bracket <> denotes averaging over the distinct elements located in a "bin" layer with a thickness of 1 nm. HOF quantifies the extent of the orientation of CNT segments with respect to the xy plane and ranges from -0.5 to 1, where the value of -0.5 corresponds to perfect in-plane alignment. It can be seen that the HOF across the relaxed "FA" and "FB" thicknesses is not reaching the -0.5 value and in fact remains slightly above HOV of the computer-generated samples. The <HOF> and <$\varphi$> over the whole relaxed samples are listed in the 5th and 6th columns of Table 2.

The film densification during relaxation can be analyzed in the density profiles shown in Figures 8c and d. Although "FA" and "FB" have the same volume filling factor, spontaneous densification is more pronounced in "FB", which lacks individual CNT dendritic branches. This result already evidences the important role played by the individual CNT branches in maintaining the sparse film structure.

Based on the density profiles shown in Figures 8c and d, we defined the film surface at the location where density becomes 0.01 g/cm³. In the 7th and 8th columns of Table 2 we report the measured relaxed film thickness $h_0$ and the density values after relaxation ($\rho_o$). The $\rho_o$ values are in the experimental range, and far different from the theoretical 0.9 g/cm³ closed packed value. In addition to the limiting bundle-size cutoff, $\varphi_{max}$ is another effective control parameter for the film structure. Confirming the direct impact of entanglement onto $h_0$, films "FC", "FD", and "FE" with larger $\varphi_{max}$ but the same volume filling factor, settle to $h_0$ values ranging from 127 nm to 209 nm, and densities $\rho_o$ ranging from 0.13 g/cm³ to 0.08 g/cm³. As a general trend, we observe that denser films are associated with the lower $E_{vdW}$ values.

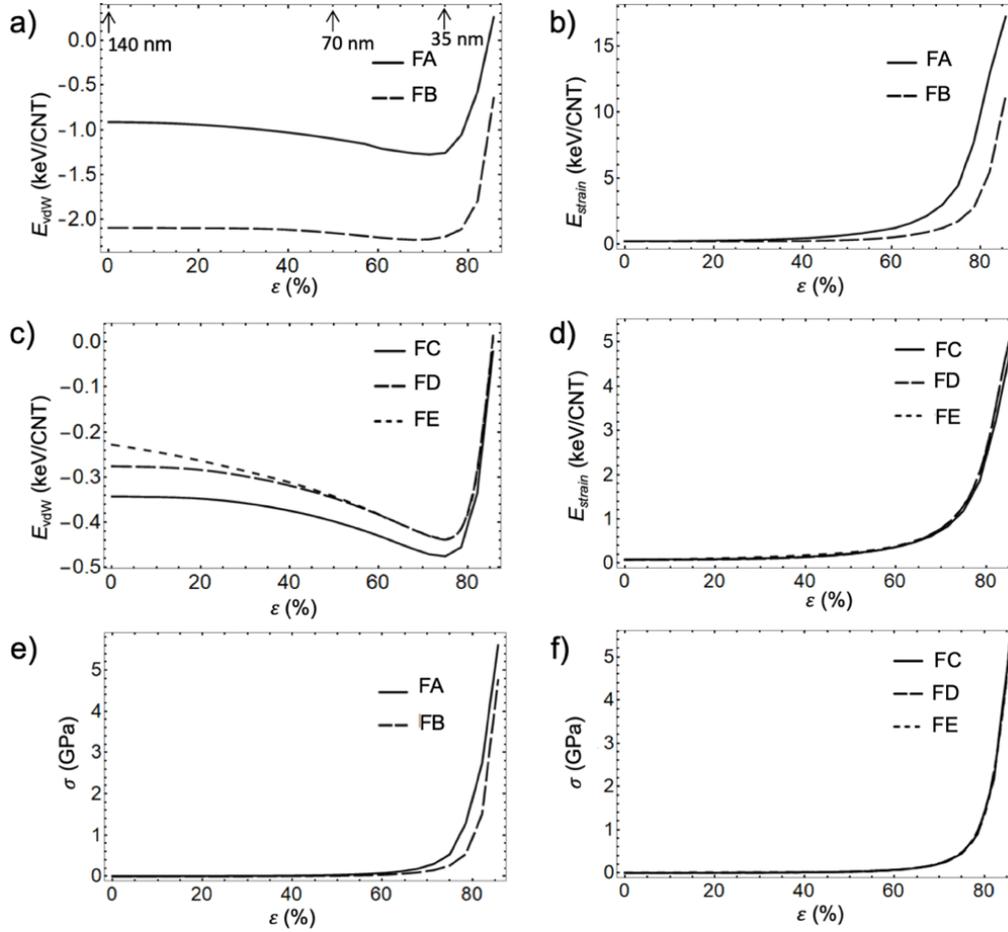

**Figure 9.** mDEM compression simulations: a) vdW energy and b) strain energy versus strain in samples "FA" and "FB". c) vdW energy and d) strain energy versus strain in samples "FC", "FD", and "FB". Stress-strain curves in samples e) "FA" and "FB", and f) "FC", "FD", and "FB".

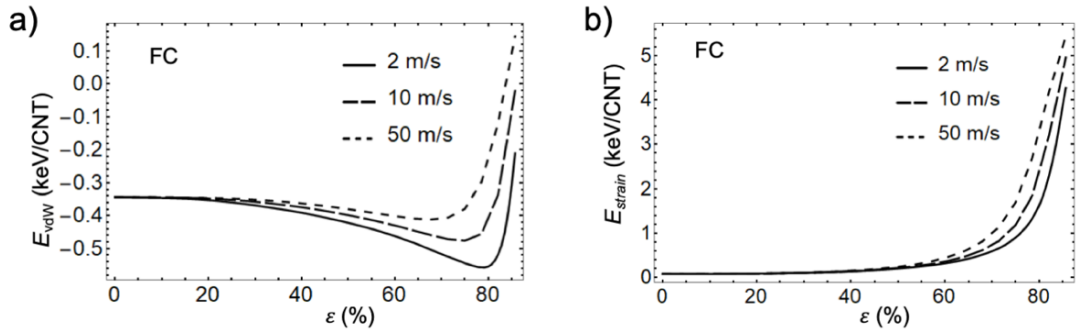

**Figure 10.** mDEM compression simulations: a) vdW energy and b) strain energy versus strain in samples "FC" at three different compression rates.

*4.5 mDEM simulations of compression-driven densification of "pristine" CNT Samples*

In the experiment, "pristine" films are further densified as they undergo compression by surface tension. To understand the process, densification under film compression was simulated by evolving the CNT film under two planar plates interacting with the film via a repulsive potential adopted from Ref.[28]. The plates are infinitely rigid and are applied to the entire two surfaces of the sample. The compressive strain ε is measured with reference to the 140 nm height of the film. Thus, the maximum compression strain of 85% considered in our simulations corresponds to a film thickness of 21 nm. The stress (σ) is defined as the sum of forces exerted by the indenter onto the interacting distinct elements divided by the surface area.

Figure 9 summarizes the results of the film compression simulations with the indenter progressing with 10 m/s. The behavior of $E_{vdW}$ and $E_{strain}$ as well as stress vs. ε, reveal two distinct deformation regimes: ultra-soft, for ε<~75%, and hardening for ε>~75%

In the ultra-soft regime, the film densification by surface compression is practically an isoenergetic process, in which the slow drop in $E_{vdW}$, Figures 9a and c, is accompanied by the slow monotonic increase of $E_{strain}$, Figures 9b and d. Further, the compression of films "FC", "FD", and "FE" show that the separation between the two regimes, which is defined by the minimum of $E_{vdW}$, depends very little on $\varphi_{max}$, which controls the initial degree of bundling in these films. The threshold strain separating the two regime exhibits a weak dependence on the applied strain rate. Figures 10a and b quantify $E_{vdW}$ and $E_{strain}$ vs. ε for sample "FC" compressed with different indenter velocities. The slowest 2 m/s case is linked to lowest $E_{vdW}$ and $E_{strain}$ curves, and to the shifting of the $E_{vdW}$ minima to about 80%.

Remarkably, Figures 9e and f, show that the stress required to compress these sample films is small (under 1 GPa) for all the considered samples, a result that agrees with the observed ability of the experimental films to densify under surface forces. In the second regime, both $E_{vdW}$ and $E_{strain}$ are dramatically increasing for all films, while the external stress required to achieve these compression levels reaches few GPa. These larger stresses values make the hardening regime likely out of reach with the surface tension compressive forces of the experimental method.

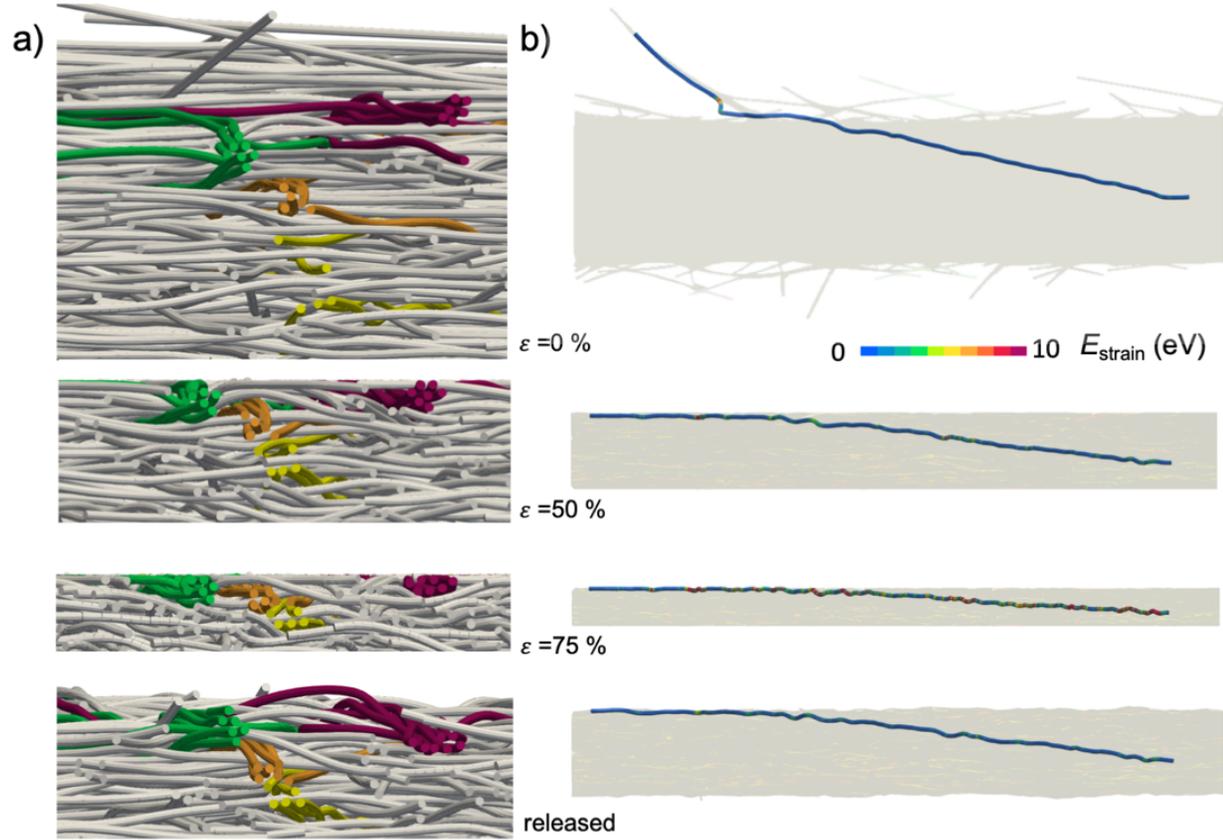

**Figure 11.** mDEM compression simulations: a) A region located near the top surface and b) a CNT of sample "FA" at different stages of compression and in the recovered film, as indicated for each frame. In a) the color is used to distinguish the evolution of existing bundles (green, orange, maroon,) and *formation of a new bundle* (yellow) and *bundle-bundle zipping* (yellow and orange). In b) the color gives the bending strain energy in the EVM bonds. The background is a portion of the substrate illustrating the shown CNT location in the film.

At the microstructural level, the applied compression alters the entanglement established in the original low-density film, opening new possibilities for CNTs to restructure under the action of vdW forces. In this respect, the simulation snapshots presented in Figure 11a evidence two $E_{vdW}$ lowering processes identified in the sample "FA":

i. At ε=50%, we observe the presence of a *new bundle,* formed by zipping of the individual CNTs depicted in yellow. The applied compression brought close enough these entangled CNTs to trigger their zipping. The inherent bending introduced by such zipping events explains the initial increase in $E_{strain}$, as quantified in Figure 9b. Note that at these higher densities, the bundles are mostly flattened.

ii. Further, by ε=75%, the new bundle zipped with an existing bundle shown in orange. The *bundle-bundle zipping* can be also inferred from the TEM image of Figure 3f.

By defining the available volume that can accommodate CNT rearrangements, density becomes an important control parameter for zipping. As density increases uniformly under compression, zipping is hindered and the response is dominated by entanglement. An important response mode of the microstructure is the development of nm-scale waviness under large ε. This is evidenced in Figure 11b, where a CNT of initial 11° inclination (before relaxation), remains tilted during compression but acquires significant bending. The bending deformation, which contributes to the increase in $E_{strain}$, is not caused by zipping but by the transversal loads exerted onto CNTs at the junctions through repulsive volume interactions.

We recall that over a few μm scale, the pristine film presents bundles that are thicker than the ones considered here. The compression response mechanisms simulated at the sub μm scale, projects that similar low stresses values will be required to compress the larger scales as long as the thicker bundles have an in-plane orientation and thus a bending dominated response (For the bundles with $\varphi_{max}>45°$ we expect an uniaxial compression dominated response.[28])

In the hardening regime, the repulsive volume interactions overcome the attractive ones to the extent that $E_{vdW}$ becomes positive at the largest considered strains, Figures 9a and c. The sharp increase in stress observed in this regime, Figures 9e and f, originates in the resistance to interpenetration of the mDEM-represented CNTs.

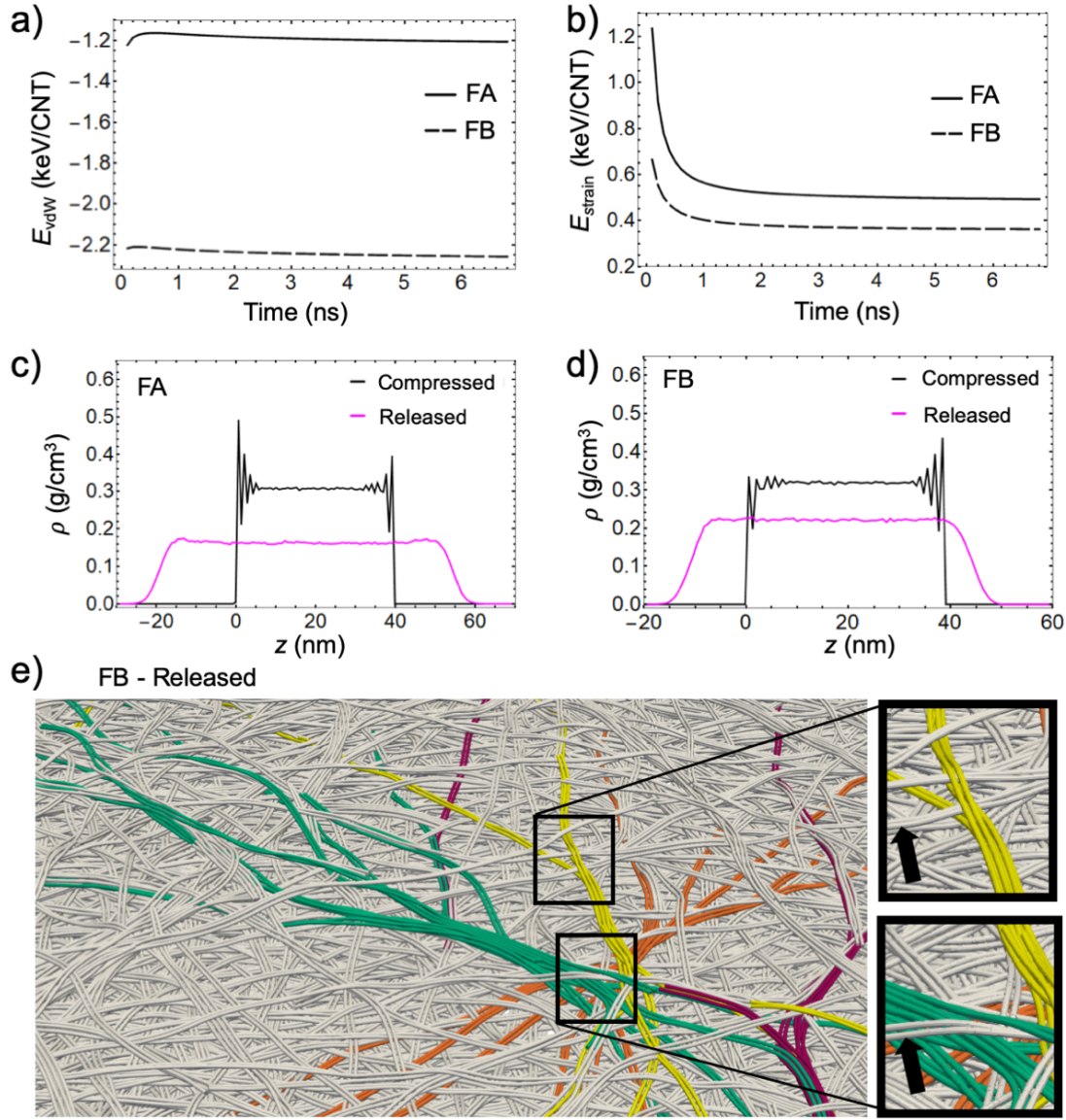

**Figure 12.** mDEM release stage simulations: a) vdW energy and b) strain energy versus time in samples "FA" and "FB". Density $\rho$ across z in c) the "FA" and d) "FB" sample before and after relaxation. e) Birds' eye view of sample "FB" after relaxation. The callouts shows examples of 2-CNT bundles (indicated by arrows) at the junctions that help maintain the branching structure.

*4.3. mDEM models of free-standing densified CNT Films*

The last set of simulations considered the release of the loads applied to the pre-compressed sample. Figures 11a and b presents the evolution of $E_{vdW}$ and $E_{strain}$ for "FA" and "FB", both pre-compressed to a $h_c$=40 nm thickness (ultra-soft regime.) Unlike in the relaxation of the computer-generated samples (Subsection 4.4), here the relaxation dynamics is driven by the lowering of $E_{strain}$. During the first 250 ps, the samples undergo spontaneous thickening with a significant release of $E_{strain}$. The initial increase in $E_{vdW}$ reflects the increase in spacing between CNTs rather than a de-bundling effect. After this initial stage, the increase in thickness slows down as the film evolves to lower both $E_{vdW}$ and $E_{strain}$ energy terms. By increasing the film thickness, the inclined tubes are lowering their waviness, as it can be seen in the last simulation snapshot of Figure 11b.

Figures 12 c and d show the mass density profiles in the compressed and released states for both "FA" and "FB" samples. Similar simulations considering the compression and bending buckling[48] of vertically aligned CNTs, reported the formation of a top dense layer near the indenter plate.[28] Our simulations show that when the constituent CNTs have low inclination angles, this surface effect becomes negligible. Remarkably, we find that the density of the compressed film is uniform over the bulk part of the film and remains uniform after release, while the surface density develops a smooth drop over a ~10 nm distance. In good agreement with the experiment, the sample "FA" ("FB") settles to a thickness value $h_d$ which is 43% (42%) smaller than the $h_o$ of the pristine form.

The parameters of all the released densified samples, including $E_{vdW}$ and $E_{strain}$, are summarized in Table 3. It is interesting to note that the total energies of the released films are very close to the ones of the corresponding pristine films, which were listed in Table 2. Thus, there is no thermodynamic drive for the densification process to occur naturally. However, the $E_{vdW}$ and $E_{strain}$ energy components are differently distributed. Reflecting the irreversible coarsening of the bundles during compression and the partial waviness stored by entangled CNTs, all densified films have lower $E_{vdW}$ energies but larger $E_{strain}$.

While the densified films models presented here are stabilized to a greater extent by adhesion, the dendritic branches model still holds, and thus the film structure is still far from a cellular one. For example, Figure 12e shows that the "FB" densified microstructure maintains the branching pattern. A small 4-CNT "small" bundle (shown in yellow) and a 16-CNT "medium" bundle (shown in green) split into thinner bundles with the help of the repulsive volume interactions with the bundles crossing. Additionally, the averaged HOF and φ listed in Table 3 are showing that the densified films still contain CNTs that are held inclined by their entanglement.

The pre-compression level is an important parameter for densification. Larger excursions into the ultra-soft regime and above it lower irreversibly $E_{vdW}$ but also introduce more bending strain, which nevertheless can be significantly lowered upon release. In agreement with this picture, our simulations show that larger pre-compression levels result, upon release, into better densified and lower energy films. For example, the data presented in the first lines of Table 3 indicated film "FA" with $h_c$=25 nm (40 nm) gives $h_d$=75.6 nm (80 nm) thickness and -0.73 keV/CNT (-0.71 keV/CNT) total energy upon release.

The angle of CNT inclination $\varphi_{max}$ is unknown experimentally. As we have seen in the pristine sample models, this parameter impacts the initial film thickness and density. To check the robustness of our results we have also investigated the free evolution of samples "FC", "FD" and "FE" after being pre-compressed at $h_c$=40 nm. Because in their pristine forms, these model films are less energetically stable than the "FA" one, they are more susceptible to densification under compression. As can be seen from the data listed in the last three lines of Table 3, the $h_d$ values are smaller than those of the "FA" sample released from the same $h_c$ values. The three thicknesses are within 2 nm, and therefore $\varphi_{max}$ appears not to be a key parameter, as long as $\varphi_{max}$ <45º.

**Table 3**. Parameters and statistical information of CNT samples after recovery from thickness $h_c$. Film thickness $h_d$, density $\rho_d$, magnitude of the Herman orientation factor <HOF>, average tilt <φ> of elements, van der Waals $E_{vdW}$, strain energy $E_{strain}$, and total energy $E$.

| Film | $h_c$ (nm) | $h_d$ (nm) | $\rho_d$ (g/cm³) | <HOF> | <φ> (°) | $E_{vdW}$ (keV/CNT) | $E_{strain}$ (keV/CNT) | $E$ (keV/CNT) |
|---|---|---|---|---|---|---|---|---|
| FA | 25 | 75.6 | 0.22 | -0.48 | 5.7 | -1.31 | 0.58 | -0.73 |
| FA | 40 | 80.0 | 0.21 | -0.48 | 5.4 | -1.20 | 0.49 | -0.71 |
| FB | 40 | 62.5 | 0.27 | -0.48 | 4.9 | -2.26 | 0.36 | -1.90 |
| FC | 40 | 74.9 | 0.22 | -0.47 | 6.1 | -0.46 | 0.19 | -0.26 |
| FD | 40 | 75.6 | 0.22 | -0.47 | 6.7 | -0.40 | 0.18 | -0.22 |
| FE | 40 | 76.3 | 0.22 | -0.45 | 8.3 | -0.38 | 0.20 | -0.19 |

Comparing the evolution of samples "FA" and "FB" both pre-compressed at $h_c$=40 nm, we see that the $h_d$ and $\rho_d$ values are not uniquely defined by $h_c$: while sample "FB" evolves toward a 62.5 nm thickness, sample "FA" settles instead to 80 nm thickness. This result clearly evidences the hindering role of the individual CNT branches present in the experimental film; achieving their restructuring by zipping under compression, as evidenced for example in Figure 11a, represents an effective mechanism for the film densification. Because of the monodisperse construction manner of sample "FA" (as opposed to the two-step formation of the pristine films discussed above) and the inherently large applied strain rate used in the simulation (as opposed to slower densification process by liquid evaporation), the densification levels obtained for "FA" should be viewed as lower limits. The densified "FB" sample with no individual CNT branches emerges as our best representation of the densified film characterized in Section 2.

5. **Summary and Conclusions**

Motivated by the advances in CNT simulation, synthesis, processing, and characterization, we utilized mesoscale simulations, electron microscopy, and spectroscopic ellipsometry to understand and characterize the densification by volatile liquid drop-casting. SEM and TEM characterization indicated that the pristine films exhibit multiscale structural features comprising CNT bundles down to the single CNTs, and suggested a coarsening of these bundles in the densified films. The spectroscopic ellipsometry measurements in transmission (pristine film) and reflection (densified film) revealed a nearly 50% change in the film thickness during ethanol evaporation.

To understand the microstructural changes caused under the surface tension, we simulated the mechanical response of the sub-$\mu$m scale to extensive compressive deformations on an ensemble of computer-generated samples with different cohesive energies but with common density and thickness values. We employed mDEM, an accurate coarse-grained simulation method which describes the microstructure of the pristine and densified CNT films as a collection of interacting dendritic bundled CNTs.

The mDEM simulations reveal significant structural changes occurring at the 100 nm scale. Compression simulations indicated that these samples undergo homogeneous densification in a two-stage regime comprising a sub-GPa plastic compression followed, at above ~75% strain, by a sharp rise of stress. In the ultra-soft regime, the film is lowering irreversibly its van der Waals cohesive energy by zipping, while increasing the bending energy of those CNTs joining the bundles but also of other entangled CNTs, which develop nm-scale waviness. Reflecting the resistance to interpenetration of the mDEM-represented CNTs, the application of few GPa stress values are required to advance into the second deformation regime. Accessing this regime is beyond reach for the surface tension stresses developed during ethanol evaporation. Thus, our simulations delineate the densification limit of the liquid processing method. Note that compression in the hardening regime it is not studied here in much detail. As thin-walled structures, small diameter CNTs may undergo a transformation from circular to hexagonal cross-section at about 1.5–1.7 GPa.[49,50] Therefore, our rigid distinct element model is not fully valid in this regime.

When removing the compressive loads, the pre-compressed samples are partially releasing the bending strain associated with the nm-scale waviness and evolve into densified films with thickness values depending on both the pre-compressed strain and the sample micro-structure. Compared to the original pristine films, the free-standing densified films have lower van der Waals adhesion but larger bending strains. Although the considered computational samples do not include dendritic bundling above the $\mu$m scale, the recovery simulations already obtained good agreement with the experiment with respect to the changes in film thickness. This indicates that the structural restructuring occurring at the 100 nm scale are responsible to a great extent for the densification process. The mesoscopic models for pristine and densified films obtained here can be further use to investigate the changes in mechanical, electrical, and optical properties driven by densification.

We emphasize that the dendritic microstructure of the pristine films proposed by our mDEM model, differs qualitatively from the usual adhesion dominated cellular structure obtained by the simpler B&S mesoscopic model. Our simulations indicate that accounting for the thinnest branches and individual CNTs is important for describing the densification process. Unfortunately, these microstructural features are not captured by the popular B&S model for CNTs. Since the realistic description of the microstructure ties directly to the mesoscale interactions, our work points to the need for accuracy in the coarse-grained modeling of CNT and of semiflexible polymer self-assembled networks[22] in general.

**Acknowledgements**

Authors thank Mr. Boris Zabelich for partial help with TEM imaging and acknowledge the OtaNano – Nanomicroscopy Center of Aalto University for a part of this research. This work was supported by NASA NNX16AE03G and by the University of Minnesota MnDrive and Grant-in-Aid programs. I.O. acknowledges the financial support from the Russian Foundation for Basic

Research (RFBR) grant 18-18-29-19198. Y.G. acknowledges the financial support from RFBR grant 18-29-20032. A. G. and A. G. N. acknowledge RFBR grant 19-32-90143. A.P.Ts. acknowledges the EDUFI Fellowship (No. TM-19-11079) from the Finnish National Agency for Education and the Magnus Ehrnrooth Foundation (the Finnish Society of Sciences and Letters) for personal financial support. T.D. greatly acknowledges support from the Fulbright U.S. Scholars program.**Data Availability**

The data that support the findings of this study are available from the corresponding author upon reasonable request.

## References

[1] E. K. Hobbie, D. O. Simien, J. A. Fagan, J. Y. Huh, J. Y. Chung, S. D. Hudson, J. Obrzut, J. F. Douglas, and C. M. Stafford, Physical Review Letters **104,** 125505 (2010).

[2] V. M. Gubarev, V. Y. Yakovlev, M. G. Sertsu, O. F. Yakushev, V. M. Krivtsun, Y. G. Gladush, I. A. Ostanin, A. Sokolov, E. Schafers, V. V. Medvedev, and A. G. Nasibulin, Carbon **155,** 734 (2019).

[3] G. A. Ermolaev, A. P. Tsapenko, V. S. Volkov, A. S. Anisimov, Y. G. Gladush, and A. G. Nasibulin, Applied Physics Letters **116,** 231103 (2020).

[4] W. Tan, J. C. Stallard, F. R. Smail, A. M. Boies, and N. A. Fleck, Carbon **150,** 489 (2019).

[5] A. M. Boies, C. Hoecker, A. Bhalerao, N. Kateris, J. de la Verpilliere, B. Graves, and F. Smail, Small **15,** 1900520 (2019).

[6] R. D. Downes, A. Hao, J. G. Park, Y. F. Su, R. Liang, B. D. Jensen, E. J. Siochi, and K. E. Wise, Carbon **93,** 953 (2015).

[7] M. Zhang, K. R. Atkinson, and R. H. Baughman, Science **306,** 1358 (2004).

8   K. Liu, Y. H. Sun, R. F. Zhou, H. Y. Zhu, J. P. Wang, L. Liu, S. S. Fan, and K. L. Jiang, Nanotechnology **21,** 045708 (2010).

9   X. P. Yu, X. H. Zhang, J. Y. Zou, Z. Y. Lan, C. Y. Jiang, J. N. Zhao, D. S. Zhang, M. H. Miao, and Q. W. Li, Advanced Materials Interfaces **3,** 1600352 (2016).

10  M. F. L. De Volder, S. J. Park, S. H. Tawfick, D. O. Vidaud, and A. J. Hart, Journal of Micromechanics and Microengineering **21,** 045033 (2011).

11  O. Yaglioglu, A. Y. Cao, A. J. Hart, R. Martens, and A. H. Slocum, Advanced Functional Materials **22,** 5028 (2012).

12  Q. Cao and J. A. Rogers, Advanced Materials **21,** 29 (2009).

13  Y. Wang, G. Drozdov, E. K. Hobbie, and T. Dumitrica, ACS Appl Mater Interfaces **9,** 13611 (2017).

14  J. Lehman, A. Sanders, L. Hanssen, B. Wilthan, J. A. Zeng, and C. Jensen, Nano Letters **10,** 3261 (2010).

15  B. D. Wood, J. S. Dyer, V. A. Thurgood, N. A. Tomlin, J. H. Lehman, and T. C. Shen, Journal of Applied Physics **118,** 013106 (2015).

16  I. Y. Stein, D. J. Lewis, and B. L. Wardle, Nanoscale **7,** 19426 (2015).

17  S. Plimpton, Journal of Computational Physics **117,** 1 (1995).

18  M. J. Buehler, Journal of Materials Research **21,** 2855 (2006).

19  A. N. Volkov and L. V. Zhigilei, Acs Nano **4,** 6187 (2010).

20  I. Ostanin, R. Ballarini, D. Potyondy, and T. Dumitrica, Journal of the Mechanics and Physics of Solids **61,** 762 (2013).

21  N. Kateris, P. A. Kloza, R. L. Qiao, J. A. Elliott, and A. Boies, Journal of Physical Chemistry C **124,** 8359 (2020).

22  E. P. DeBenedicts, Y. Zhang, and S. Keten, Macromolecules **53,** 6123 (2020).

23  L. V. Zhigilei, C. Wei, and D. Srivastava, Physical Review B **71** (2005).

24  I. Ostanin, R. Ballarini, and T. Dumitrica, Journal of Applied Mechanics-Transactions of the ASME **81,** 061004 (2014).

25  A. N. Volkov and L. V. Zhigilei, Journal of Physical Chemistry C **114,** 5513 (2010).

26  Y. Z. Wang, H. Xu, G. Drozdov, and T. Dumitrica, Carbon **139,** 94 (2018).

27  B. K. Wittmaack, A. N. Volkov, and L. V. Zhigilei, Composites Science and Technology **166,** 66 (2018).

28  B. K. Wittmaack, A. N. Volkov, and L. V. Zhigilei, Carbon **143,** 587 (2019).

29  M. Alzaid, J. Roth, Y. Z. Wang, E. Almutairi, S. L. Brown, T. Dumitrica, and E. K. Hobbie, Langmuir **33,** 7889 (2017).

30  I. Ostanin, T. Dumitrica, S. Eibl, and U. Rude, Journal of Applied Mechanics-Transactions of the ASME **86,** 121006 (2019).

31  I. Itasca Consulting Group,  PFC3D Particle Flow in 3 Dimensions 5.00. (2016).

32  I. E. Berinskii, A. Y. Panchenko, and E. A. Podolskaya, Modelling and Simulation in Materials Science and Engineering **24,** 045003 (2016).

33  I. A. Ostanin, P. Zhilyaev, V. Petrov, T. Dumitrica, S. Eibl, U. Ruede, and V. A. Kuzkin, Letters on Materials-Pis Ma O Materialakh **8,** 240 (2018).

34  T. Preclik and U. Rude, Computational Particle Mechanics **2,** 173 (2015).

35  V. A. Kuzkin and I. E. Asonov, Physical Review E **86,** 051301 (2012).

36  T. Anderson, E. Akatyeva, I. Nikiforov, D. Potyondy, R. Ballarini, and T. Dumitrică, Journal of Nanotechnology in Engineering and Medicine **1,** 041009 (2010).

37  H. Xu, J. Al-Ghalith, and T. Dumitrica, Carbon **134,** 531 (2018).

[38] S. M. Johnson, J. R. Williams, and B. K. Cook, International Journal for Numerical Methods in Engineering **74,** 1303 (2008).

[39] Y. Tian, M. Y. Timmermans, M. Partanen, A. G. Nasibulin, H. Jiang, Z. Zhu, and E. I. Kauppinen, Carbon **49,** 4636 (2011).

[40] O. Reynaud, A. G. Nasibulin, A. S. Anisimov, I. V. Anoshkin, H. Jiang, and E. I. Kauppinen, Chemical Engineering Journal **255,** 134 (2014).

[41] D. Satco, D. S. Kopylova, F. S. Fedorov, T. Kallio, R. Saito, and A. G. Nasibulin, Acs Applied Electronic Materials **2,** 195 (2020).

[42] J. N. Hilfiker, B. Pietz, B. Dodge, J. N. Sun, N. N. Hong, and S. Schoeche, Applied Surface Science **421,** 500 (2017).

[43] N. C. Passler and A. Paarmann, Journal of the Optical Society of America B-Optical Physics **34,** 2128 (2017).

[44] K. Mustonen, T. Susi, A. Kaskela, P. Laiho, Y. Tian, A. G. Nasibulin, and E. I. Kauppinen, Beilstein Journal of Nanotechnology **3,** 692 (2012).

[45] Y. Li and M. Kroger, Applied Physics Letters **100,** 021907 (2012).

[46] I. Zacharov, R. Arslanov, M. Gunin, D. Stefonishin, A. Bykov, S. Pavlov, O. Panarin, A. Maliutin, S. Rykovanov, and M. Fedorov, Open Engineering **9,** 512 (2019).

[47] Y. Z. Wang, M. R. Semler, I. Ostanin, E. K. Hobbie, and T. Dumitrica, Soft Matter **10,** 8635 (2014).

[48] I. Nikiforov, D. B. Zhang, R. D. James, and T. Dumitrica, Applied Physics Letters **96,** 123107 (2010).

[49] C. Onuoha, G. Drozdov, Z. Y. Liang, G. M. Odegard, E. J. Siochi, and T. Dumitrica, ACS Applied Nano Materials **3,** 5014 (2020).

[50] M. J. Lopez, A. Rubio, J. A. Alonso, L. C. Qin, and S. Iijima, Physical Review Letters **86,** 3056 (2001).